\documentclass[prl,twocolumn,showpacs,superscriptaddress,preprintnumbers]{revtex4-2}
\usepackage{amsmath,amssymb,amsfonts,float,graphics,epsfig,epstopdf,color,verbatim,tabularx,bm,multirow,appendix}
\usepackage{tikz}
\usepackage{amsmath,amssymb,graphicx}
\usepackage{wasysym}
\usepackage[utf8]{inputenc}
\usepackage[T1]{fontenc}
\usepackage{xcolor}
\usepackage{dsfont}
\usepackage{textcomp}
\usepackage{bm}

\usepackage{graphicx}%
\usepackage{xcolor}
\usepackage{dcolumn}
\usepackage{bm}
\usepackage{xspace}
\usepackage{time}
\usepackage{booktabs}
\usepackage{multirow}
\usepackage{makecell}

\pdfoutput=1
\usepackage{color}
\definecolor{LinkColor}{rgb}{0.256,0.439,0.588}
\usepackage{hyperref}
\hypersetup{
colorlinks=true,
citecolor=LinkColor,
linkcolor=LinkColor,
urlcolor=LinkColor
}

\newcommand{\newsect}[1]{\noindent \textit{\textcolor{blue}{#1.-}}}

\newcommand{\bra}[1]{\left \langle#1\right \rvert}
\newcommand{\ket}[1]{\left \lvert#1\right \rangle}

\newcommand{\EPJJ}{\mathrm{EPJ}_1 \mathrm{J}_2}

\newcommand{\lshape}{%
    \begin{tikzpicture}[scale=0.5, baseline=-0.5ex]
        \coordinate (A) at (0, -0.5);
        \coordinate (B) at (1, -0.5);
        \coordinate (C) at (0, 0.5);
        \draw[thick] (A) -- (B);
        \draw[thick] (A) -- (C);
        \foreach \point in {A, B, C}
            \filldraw (\point) circle (3pt);
    \end{tikzpicture}%
}
\newcommand{\plaquette}{%
    \begin{tikzpicture}[scale=0.5, baseline=-0.5ex]
        \coordinate (A) at (0, -0.5);
        \coordinate (B) at (1, -0.5);
        \coordinate (C) at (1, 0.5);
        \coordinate (D) at (0, 0.5);
        \draw[thick] (A) -- (B) -- (C) -- (D) -- cycle;
        \foreach \point in {A, B, C, D}
            \filldraw (\point) circle (3pt);
    \end{tikzpicture}%
}
\newcommand{\hexagonwithdots}{%
    \begin{tikzpicture}[scale=0.4, baseline=-0.5ex] 
        \coordinate (A) at (0:1);
        \coordinate (B) at (60:1);
        \coordinate (C) at (120:1);
        \coordinate (D) at (180:1);
        \coordinate (E) at (240:1);
        \coordinate (F) at (300:1);
        \draw[thick] (A) -- (B) -- (C) -- (D) -- (E) -- (F) -- cycle;
        \foreach \point in {A, B, C, D, E, F}
            \filldraw (\point) circle (3pt);
    \end{tikzpicture}%
}

\DeclareMathOperator{\Tr}{Tr}
\graphicspath{./figures}

\begin{document}

\title{Entanglement architecture of beyond-Landau quantum criticality}

\author{Menghan Song}
\affiliation{Department of Physics and HK Institute of Quantum Science \& Technology, The University of Hong Kong, Pokfulam Road, Hong Kong SAR, China}

\author{Ting-Tung Wang}
\affiliation{Department of Physics and HK Institute of Quantum Science \& Technology, The University of Hong Kong, Pokfulam Road, Hong Kong SAR, China}

\author{Liuke Lyu}
\affiliation{D\'epartement de Physique, Universit\'e de Montr\'eal, Montr\'eal, QC H3C 3J7, Canada}

\author{William Witczak-Krempa}
\affiliation{D\'epartement de Physique, Universit\'e de Montr\'eal, Montr\'eal, QC H3C 3J7, Canada}
\affiliation{
 Institut Courtois, Universit\'e de Montr\'eal, Montr\'eal (Qu\'ebec), H2V 0B3, Canada
}
\affiliation{
 Centre de Recherches Math\'ematiques, Universit\'e de Montr\'eal, Montr\'eal, QC, Canada, HC3 3J7
}

\author{Zi Yang Meng}
\affiliation{Department of Physics and HK Institute of Quantum Science \& Technology, The University of Hong Kong, Pokfulam Road, Hong Kong SAR, China}

\begin{abstract}
    Quantum critical points beyond the Landau paradigm exhibit fractionalized excitations and emergent gauge fields. 
    Here, we use entanglement microscopy--full tomography of the reduced density matrix of small subregions and subsequent extraction of their quantum correlations--to resolve the entanglement architecture near such exotic critical points. We focus on genuine multipartite entanglement (GME). Through unbiased quantum Monte Carlo sampling of RDMs across conventional O(2)/O(3) Wilson-Fisher transitions, and unconventional XY$^*$, and N\'eel-VBS transitions in (2+1)d, we discover a  dichotomy: Landau criticality amplifies GME within compact subregions, while non-Landau criticality redistributes entanglement into larger, loopy configurations. Key signatures at non-Landau criticality include the absence of three-spin GME, and the loss of non-loopy entanglement in unicursal regions. Similar results in a critical resonating valence bond wavefunction confirm this multipartite entanglement structure as a common feature of emergent gauge theories. Our findings reveal a distinct  entanglement architecture in beyond-Landau quantum critical theories.
\end{abstract}

\date{\today}

\maketitle

\newsect{Introduction}
Quantum phase transitions (QPTs), marking zero-temperature transitions between distinct phases of matter, represent fundamental reorganizations of quantum entanglement~\cite{Sachdev2011Quantum}. The Landau-Ginzburg-Wilson paradigm has long provided the framework for understanding conventional QPTs, where symmetry breaking governs critical behavior of different universality classes~\cite{landau1980statistical}. Recent decades, however, have witnessed quantum critical points (QCPs) beyond this paradigm—exhibiting fractionalized excitations, emergent gauge fields, and topological order~\cite{Senthil2004Quantum}. 
These exotic critical theories challenge traditional characterization methods such as the local order parameter. One promising direction is to probe how different parts of the system are entangled with each other.

Entanglement measures have proven to be powerful tools for studying exotic quantum phase matters and criticality. For example, entanglement entropy captures universal constants in symmetry-broken phases, critical points, and topological phases~\cite{Bohdan2015Detecting,PStoudenmire2014Corner,Bueno2015Universality,Liao2024Extracting,Kitaev2006Topological}. Non-Landau QCPs are believed to be more entangled than conventional Landau ones, inferred by an additional constant term in the entanglement entropy scaling form~\cite{Swingle2012Structure,Isakov2011Topological}.
{
However, the same amount of entanglement shared between two subregions can result from different types of interactions, which makes entanglement entropy too coarse to characterize the underlying structure. 
Fortunately, tools from quantum information that quantify genuine multipartite entanglement (GME) tackle this problem by resolving how entanglement is shared among three or more subregions.
Recent work established GME as a sharp probe of exotic phases of matter: in quantum Ising model, all forms of GME become maximal near the QCP~\cite{Lyu2025multiparty}, while in quantum spin liquids (QSL), frustrated interactions prevent nearby spins from sharing entanglement (termed ``entanglement frustration''), so GME vanishes on all minimal non-loopy clusters and appears exclusively on closed loops~\cite{lyu2025loop}.
Taken together, these observations imply sharply different reorganizations of multipartite entanglement at Landau and non-Landau QCPs, motivating a systematic study of entanglement structure across various quantum phase transitions.
}

To investigate this contrast, we apply entanglement microscopy—a tomographic approach building on quantum Monte Carlo sampling of reduced density matrices—that directly resolves multipartite entanglement structure within finite subregions~\cite{Wang2025Entanglement}. This technique enables subregion tomography.
{We then apply an entanglement monotone called genuine multipartite negativity (GMN)~\cite{Guhne2011}, to detect the entanglement between different parts of microscopic subregions at criticality.
Our results first confirm the spin-liquid ``entanglement frustration" picture with large-scale QMC on the Balents-Fisher-Girvin (BFG) spin liquid phase~\cite{balentsFractionalization2002}: GMN vanishes on non-loopy regions and survives only on loops.
{We then explore a range of phase transitions, including conventional O(2)/O(3) Wilson-Fisher transitions,  unconventional XY$^*$ and N\'eel-VBS transitions in (2+1)d.}
This reveals a fundamental dichotomy in how criticality structures multipartite entanglement: Landau QCPs enhance GMN on all connected subregions, including the minimal three-spin cluster, whereas non-Landau QCPs suppress GMN in small subregions and redistribute entanglement onto larger, loopy regions.
}


\begin{figure*}[htp!]
\includegraphics[width=2\columnwidth]{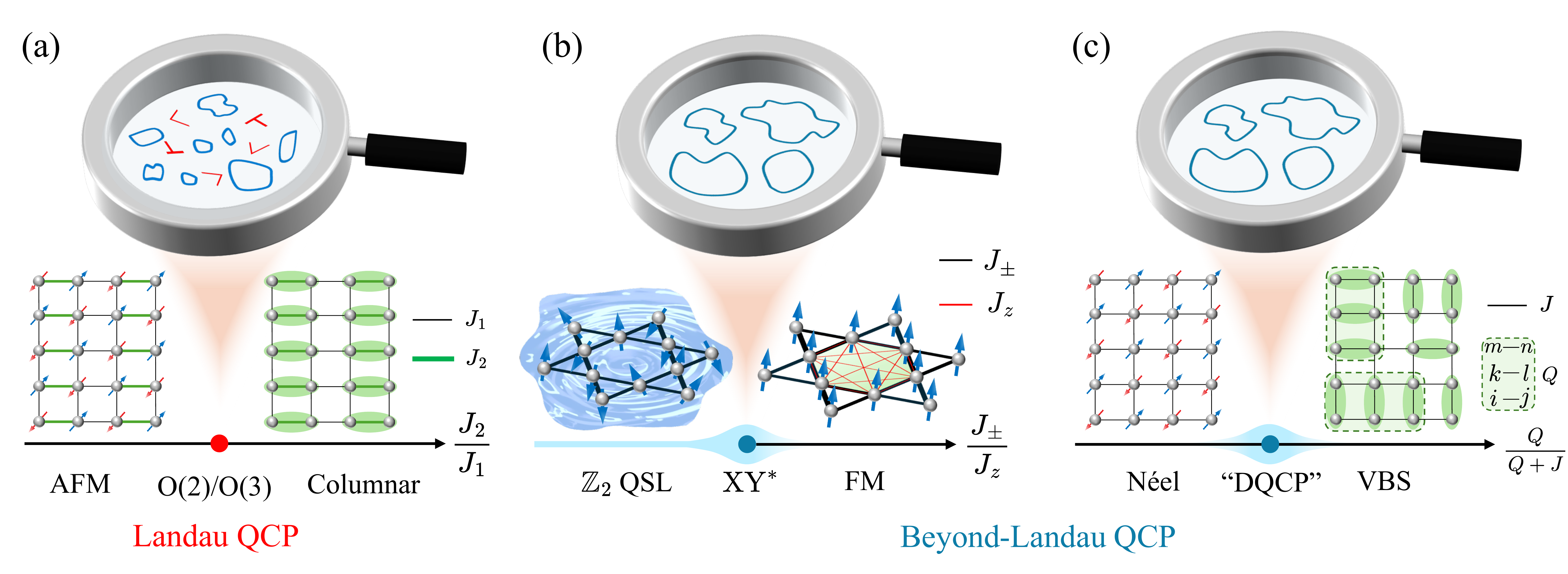}
\caption{\textbf{Phase diagrams and entanglement microscopy at Landau and beyond-Landau quantum critical points.} Panels (a), (b), and (c) demonstrate the phase diagram of $H_{\EPJJ}$,$H_{\mathrm{BFG}}$, and $H_{JQ_3}$ together with lattice structure, respectively. The phase diagrams in panels (a) (Landau) and (b)(c) (Beyond-Landau) reveal contrasting entanglement architecture: at the Landau QCP, entanglement can be found in small, non-loopy subregions (shown in the microscope), while at QCPs governed by emergent and fractionalized degrees of freedom (beyond-Landau), it primarily resides within large and loopy subregions (shown in the microscope). We use red to represent Landau criticality and blue to portray the critical points and phase with the presence of fractalization and emergent gauge fields.}
\label{fig:fig1}
\end{figure*}

\newsect{Genuine multipartite entanglement}
Consider a pure quantum state $|\psi\rangle$ defined on an $n$-party Hilbert space $\mathcal{H}_1 \otimes \cdots \otimes \mathcal{H}_n$. This state is  \textbf{fully separable} if it can be factorized as $|\psi\rangle = |\psi_1\rangle \otimes |\psi_2\rangle \otimes \cdots \otimes |\psi_n\rangle$, otherwise, it contains some form of entanglement. For an $N$-party system, a state is considered genuinely multipartite entangled if it cannot be represented as a mixture of states separable across any bipartition. For example, a three-party ($A, B, C$) state is biseparable if it can be written as $\rho^{\text{bsep}} = p_1 \rho_{A|BC}^{\text{sep}} + p_2 \rho_{B|AC}^{\text{sep}} + p_3 \rho_{C|AB}^{\text{sep}}$, where each term is separable with respect to a specific bipartition (e.g., $A|BC$). States that defy this form exhibit genuine multipartite entanglement. 
Genuine multipartite negativity is a computable entanglement monotone for GME~\cite{Guhne2011,Hofmann_2014}. We first define the minimum negativity across all bipartitions ${N}_{\text{min}}(\rho) = \min_{M|\overline{M}} {N}_{M|\overline{M}}(\rho)$, 
where ${N}_{M|\overline{M}}(\rho) = \frac{1}{2} ( \|\rho^{T_M}\|_1 - 1 )$ quantifies the negativity for a bipartition $M|\overline{M}$, with $\rho^{T_M}$ being the partial transpose over subsystem $M$ and $\|\cdot\|_1$ the trace norm. Then, GMN is defined as the mixed convex-roof extension of $N_\text{min}$ as
\begin{equation}
    \mathcal{N}(\rho)=\min _{\{p_k,\rho_k\}} \sum_k p_k \mathcal{N}_{\text{min}}\left(\rho_k\right),
    \label{eq:CRE_GMN}
\end{equation}
where each ensemble $\{p_k,\rho_k\}$ satisfies $\rho=\sum_k p_k \rho_k$. A positive GMN ($\mathcal{N}_{\text{GM}}(\rho) > 0$) provides a sufficient but not necessary condition for GME. We note that the GMN reduces to the usual negativity in the bipartite case.

\newsect{Models and Methods}
We employ the entanglement microscope technique to sample the reduced density matrices and then detect GME within microscopic regions in quantum many-body systems~\cite{Wang2025Entanglement}. We use the GMN to construct an entanglement witness and study the multiparty entanglement between $m$ parties with $m\le6$. With the sampled RDMs, GMN can be computed efficiently via a semidefinite program~\cite{Guhne2011}. Details of the minimization process to find GMN can be found in the Supplemental Material (SM)~\cite{suppl}.

We first investigate genuine multipartite entanglement at conventional Landau phase transitions, focusing on the conformal-invariant critical points of the O(2)/O(3) transition via the easy-plane $J_1-J_2$ ($\EPJJ$) Hamiltonian,
\begin{equation}
 H_{\EPJJ} = J_1 \sum_{\langle i,j \rangle} D_{ij} + J_2 \sum_{\langle i,j \rangle'} D_{ij},
\end{equation}
where $D_{ij} = S^x_i S^x_j + S^y_i S^y_j + \Delta S^z_i S^z_j$. $\langle i,j\rangle$ denotes the thin black $J_1$ bonds and $\langle i,j\rangle^\prime$ represents the thick green $J_2$ bonds, as depicted in Fig.~\ref{fig:fig1}(a). $\Delta$ introduces the easy-plane anisotropy. At $\Delta=1$, the Hamiltonian holds SU(2) spin rotational symmetry and can realize the (2+1)d O(3) QCP belonging to the Heisenberg universality class at $g\equiv J_2/J_1=1.90951(1)$~\cite{zhaoScaling2022,maAnomalous2018}.
When $\Delta \in [0,1)$, the SU(2) symmetry is explicitly broken down to U(1) and $H_{\EPJJ}$ realizes a (2+1)d O(2) Wilson-Fisher QCP in the XY universality class~\cite{MHasenbusch1999high}. We fix $\Delta=1/2$ such that the QCP  $g = 2.735(2)$~\cite{Ma2018Dynamical} separates the antiferromagnetic (AFM) and columnar phase.

For a direct comparison with the conventional Landau transitions, we study the GME across quantum phase transitions with "deconfined" degrees of freedom living at the critical point (or more generally survive in one of the phases)~\cite{Swingle2012Structure}. This type of exotic phase transition can exhibit the emergence of gauge fields and degrees of freedom that possess fractional quantum numbers. 
A simple candidate of such transitions is a QPT between a ferromagnetic (superfluid) and a $\mathbb{Z}_2$ quantum spin liquid (boson Mott insulator), also known as the XY$^\ast$ QCP (possess the same exponents $\nu$ and $z$ as the usual XY universality class, but with large anomalous dimension $\eta$~\cite{isakovUniversal2012}). It can be realized in a sign-problem-free BFG Hamiltonian on a kagome lattice~\cite{balentsFractionalization2002},
$$
H_{\mathrm{BFG}} = -J_{\pm} \sum_{\langle i,j \rangle} (S_i^+ S_j^- + \text{h.c.}) + \frac{J_z}{2} \sum_{\hexagon} \left( \sum_{i \in \hexagon} S_i^z \right)^2,
$$
as shown in Fig.~\ref{fig:fig1} (b). The BFG Hamiltonian contains both nearest-neighbor spin flip interactions $J_\pm > 0$ and a plaquette interaction $J_z > 0$ for each hexagon. The phase diagram of the model, from the $\mathbb{Z}_2$ QSL at small $J_{\pm}/J_z$ to ferromagnetic phase at large $J_{\pm}/J_z$, through a (2+1)d XY$^\ast$ QCP at$(J_{\pm}/J_z)_c=0.07076(2)$ has been well studies in the literature~\cite{isakovSpin2006,isakovUniversal2012,Isakov2011Topological,wangQuantum2018,wangFractionalized2021}. Inside the $\mathbb{Z}_2$ QSL phase, there exist fractionalised vison excitations with a small gap in the scale of $0.01\sim J_z$ and spinon excitations with a large gap in the scale of $0.4\sim J_z$~\cite{wangFractionalized2021}. 
This transition can be effectively characterized as a condensation of spinons, with the vison gap remaining finite. Notably, the deconfinement of the fractional boson has already taken place at the critical point and persists into the Mott phase, making the comparison of GME between XY$^\ast$ QCP and Landau transitions interesting.

Another example of non-Landau criticality is the transition from Néel to valence bond solid (VBS) states, which can be described by emergent bosons with fractional spin coupled to a gapless U(1) gauge field~\cite{Senthil2004Quantum}. We investigate the $JQ_3$ model on a square lattice introduced by Sandvik~\cite{Sandvik2007Evidence,Lou2009Anti} with the Hamiltonian,
\begin{equation}
H_{JQ_{3}}=-J \sum_{\langle i j\rangle} P_{i, j}-Q \sum_{\langle i j k l m n\rangle} P_{i j} P_{k l} P_{m n},
\end{equation}
where $P_{ij} = 1/4 -\mathbf{S}_i \cdot \mathbf S_j$ is a spin singlet projector and the $Q$ term is a six-spin plaquette interaction as shown in Fig.~\ref{fig:fig1}(c). The putative quantum critical point separating the AFM and VBS phases is at $q\equiv Q/(J + Q) = 0.59864(4)$.
This model was thought to realize a deconfined quantum critical point (DQCP) in (2+1)d, but recently, evidence was given in support of a weakly first-order transition~\cite{zhaoScaling2022,song2025evolution,Deng2024Diagnosing,zhaoScaling2024}. Close to the critical point, a finite order moment was detected at large system sizes ($L>300$)~\cite{takahashi2024so5multicriticalitytwodimensionalquantum}; nevertheless, at small lattice sizes, the pseudo-critical behaviour~\cite{Emidio2024properlattice,Ma2020TheoryofDeconfinedPsedo} arises, and the low-energy theory of DQCP contains strongly coupled gauge fields~\cite{Swingle2012Structure,Senthil2004Quantum,Wang2017Deconfined,Lu2021Self}. We note that the VBS phase is characterized by a fourfold degeneracy, leading to domain walls in finite-size lattices that introduce sample errors. To mitigate this uncertainty, we apply a small pinning field of $\delta/L$ at the $J_2$ bonds defined in $H_{\EPJJ}$. This adjustment lifts the fourfold degeneracy in the VBS phase while preserving the Néel phase and critical properties. With the pinning field in place, we can effectively investigate GMN for a specific VBS pattern, ensuring satisfactory data quality. (See SM~\cite{suppl} for the effects of different pinning field strengths.)

We use the stochastic series expansion algorithm~\cite{sandvik1999sse,Wang2025Entanglement, Mao2025Sampling} to sample the RDMs for the above Hamiltonians. The directed loop updating scheme~\cite{sandvik2002directed} is used for Hamiltonians without a global SU(2) symmetry.

\begin{figure}[htp!]
\includegraphics[width=\columnwidth]{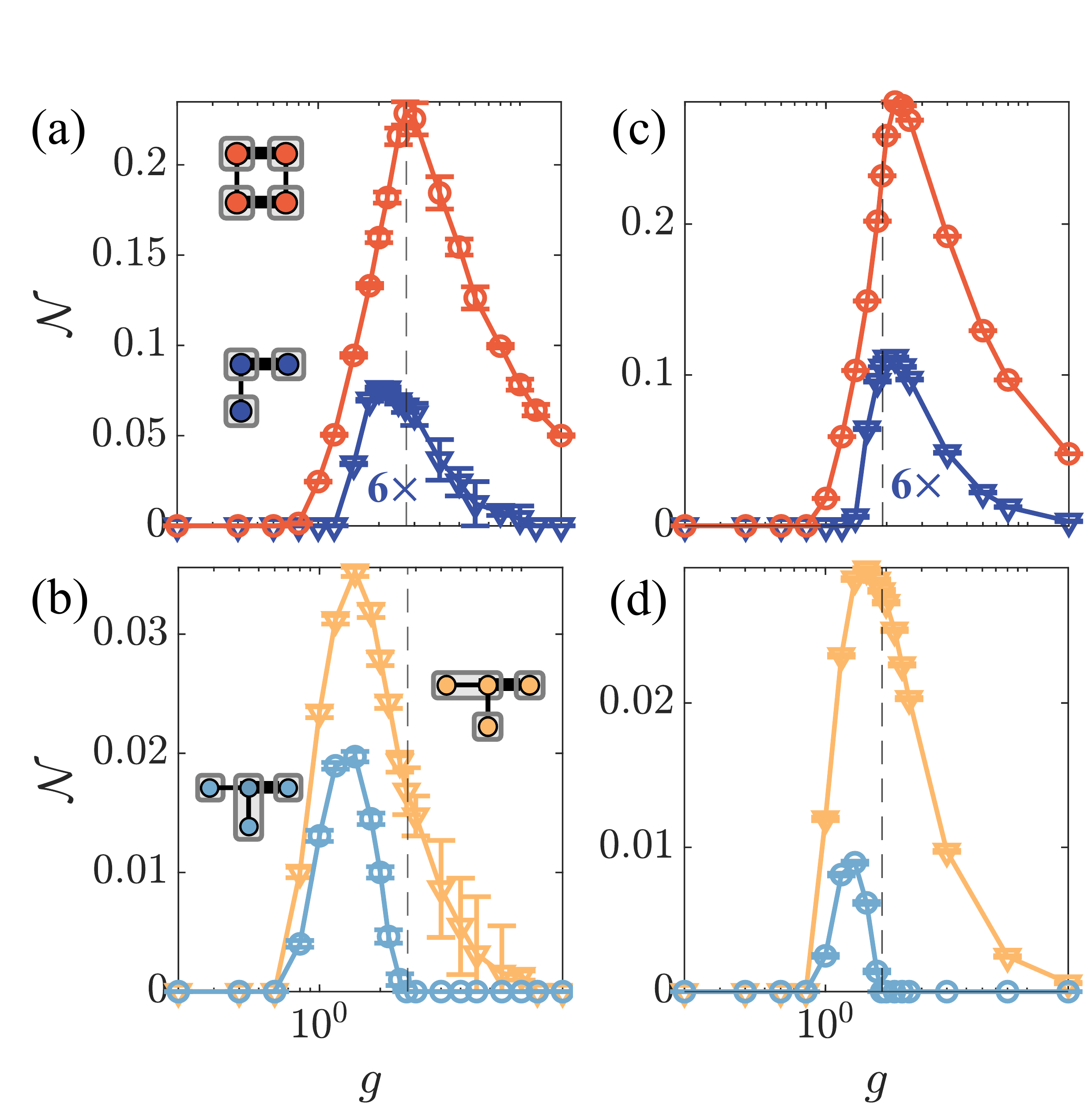}
\caption{\textbf{GMN across (2+1)d O(2) and O(3) quantum critical points.} GMN at different partitions is obtained from the $\EPJJ$ model at $\Delta=1/2$ (a)(b) and $\Delta=1$ (c)(d). QMC simulations are performed at $L=24$ and inverse temperature $\beta=48$. The GMN for each specific partition (illustrated in the diagrams) is plotted in a matching color. Some GMN values are multiplied by the factors with corresponding colors for better clarity. The dashed line indicates the location of the QCP: $g_c= 2.735(2)$ for $\Delta=1/2$ and $1.90951(1)$ for $\Delta=1$. (a) and (c) show the GMN for unicursal subregions, while (b) and (d) show that for non-unicursal ones.}
\label{fig:O2O3_GMN}
\end{figure}

\newsect{GMN at Landau and beyond-Landau QCPs}
Fig.~\ref{fig:O2O3_GMN} displays the GMN across the O(2) and O(3) QCPs, exemplifying conventional Landau phase transitions. At the limits \(g = 0\) and \(g = \infty\), \(H_{\EPJJ}\) reduces to decoupled spin ladders and a product of singlets, respectively. Both ground states are fully separable product states of the form \(\bigotimes_{i} \ket{\psi_i}\), where \(\ket{\psi_i}\) denotes the state of the \(i\)-th subregion. 
Consequently, at \(g=0\) and \(g=\infty\), separable ground states yield vanishing GMN. 
Most notably, GMN peaks near the QCPs for three-site L subregion and the four-site plaquette, but have a sudden death in the AFM phase; for the O(2) transition, the plaquette and three-site GMN vanish at $g\approx0.6$ and 1.2, while for O(2), they vanish at 0.9 and 1.2, respectively.
This critical enhancement of GMN mirrors our previous observations for Ising criticality across various dimensions~\cite{Lyu2025multiparty}. Therefore, quantum critical fluctuations in Landau QCPs enhance GME within minimal subregions despite divergent correlation lengths.

\begin{figure}[htp!]
\includegraphics[width=\columnwidth]{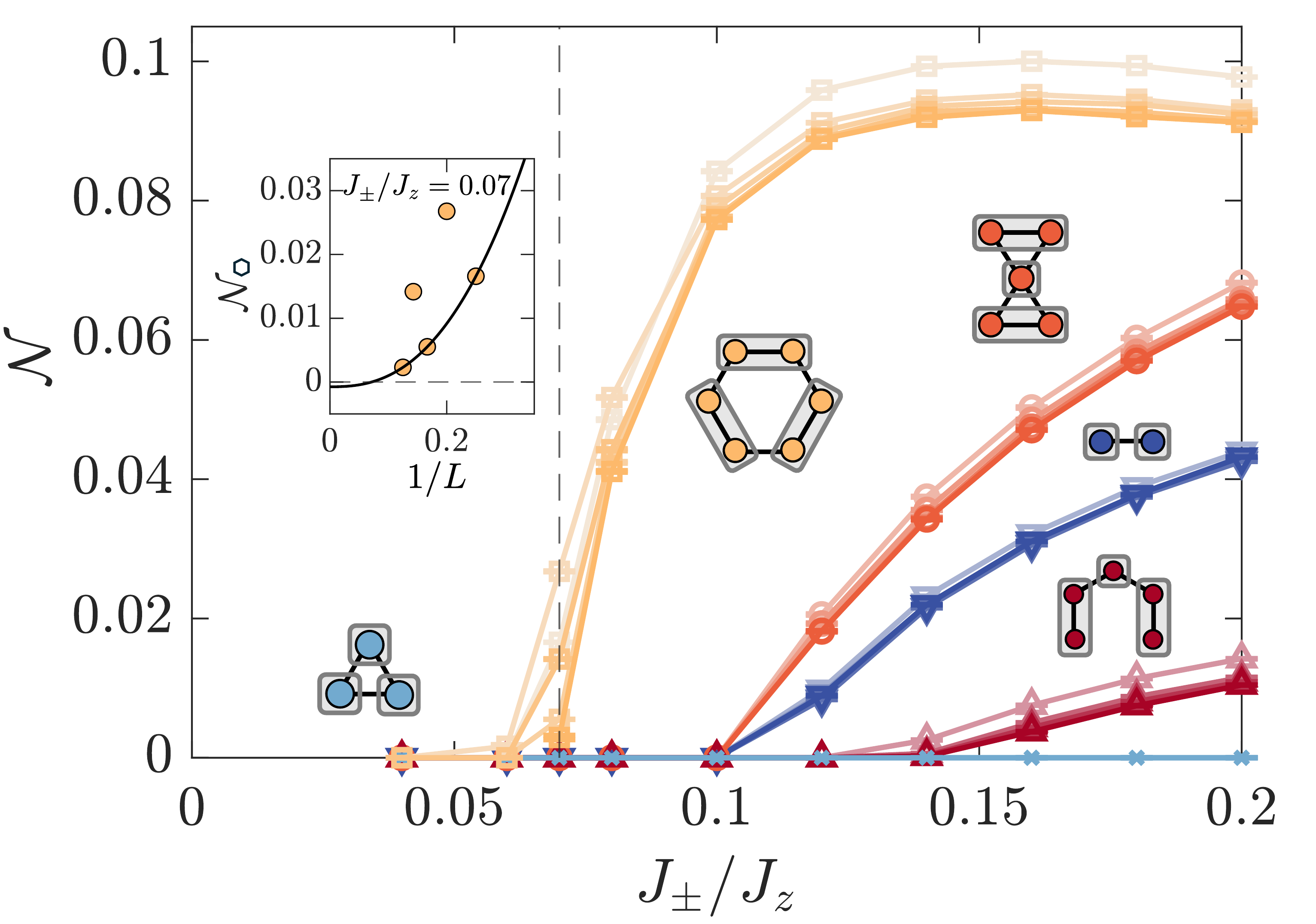}
\caption{\textbf{GMN across (2+1)d XY$^\ast$ quantum critical point.} GMN for various partitions of the BFG model in the ground-state limit (see SM~\cite{suppl} for the temperature convergence). Curve colors match their corresponding partition diagrams. The dashed line marks the QCP at $J_\pm/J_z\approx 0.07$. 
Line color intensity represents the system size, ranging from light ($L_{\mathrm{min}}=4$) to dark ($L_{\mathrm{max}}=7$, or 8 for the hexagon).
The inset shows the finite-size extrapolation of the hexagon's three-party GMN at the QCP ($L=4$ to $8$), revealing an even-odd oscillation. A power-law fit to the even-size data is shown by the black line.}
\label{fig:BFG_GMN_all_size}
\end{figure}

\begin{figure}[htp!]
\includegraphics[width=\columnwidth]{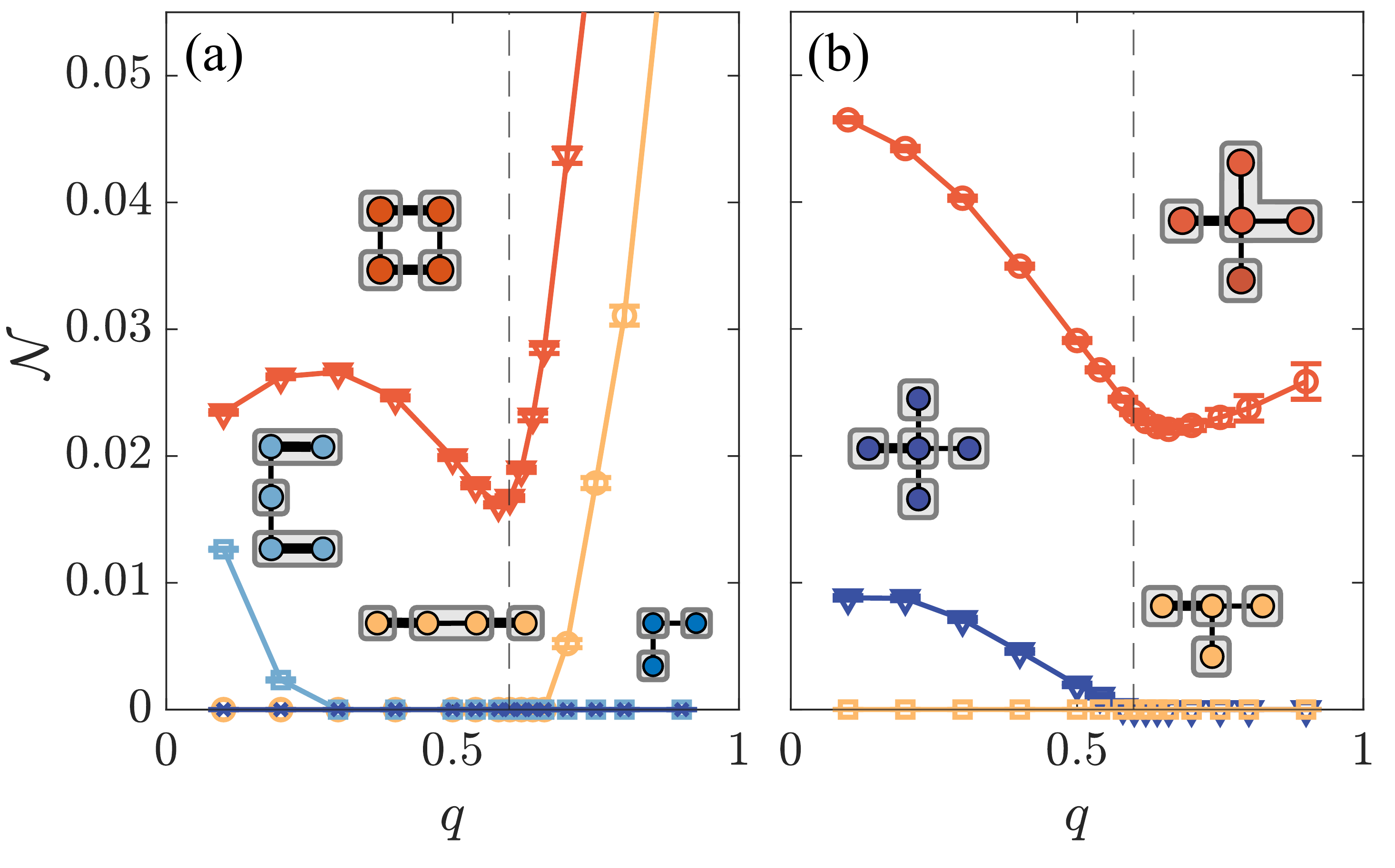}
\caption{\textbf{GMN across N\'eel-VBS weakly first-order transition of JQ model.} GMN for different partitions is obtained from the $JQ_3$ model at $L=24$, inverse temperature $\beta=96$ and pinning field $\delta=0.01$. (a) GMN for unicursal subregions. (b) GMN for non-unicursal subregions. Curve colors match their corresponding partition diagrams. Thicker bonds in the partition diagram indicate where the pinning field $\delta$ is applied. The dashed line indicates the location of the weakly first-order N\'eel-VBS transition at $q_c\approx 0.6$.}
\label{fig:jq3GMN}
\end{figure}

GMN shows different behavior at quantum critical points beyond the Landau paradigm. Figure~\ref{fig:BFG_GMN_all_size} displays GMN for the XY$^*$ transition in the BFG model. At small positive $J_\pm/J_z$, the ground state is a $\mathbb{Z}_2$ spin liquid with fractionalized excitations. In this case, criticality depends on the condensation of fractionally charged bosons (spinons). Previous studies reported strong entanglement frustration in the quantum spin liquid phases: GME disappearing in non-loopy subregions~\cite{lyu2025loop}. Consistent with this picture, we do not find any non-zero GMN values within the $\mathbb{Z}_2$ spin liquid phase at numerically accessible subregions. 
In addition, we also confirm another aspect of entanglement frustration: as we approach the QCP from the ferromagnetic phase, GMN decreases more quickly in non-loopy subregions (vertex-truncated hexagon), than in loopy configurations (bowtie and hexagon).
Importantly, all subregions except the hexagon show vanishing GMN before reaching criticality. At the QCP, the hexagon's three-party GMN shows stronger finite-size effects compared to the ferromagnetic phase. Finite-size extrapolation (inset, Fig.~\ref{fig:BFG_GMN_all_size}) with a power-law fitting ($aL^{-b} + c$) on $L=4,6,8$ data results in a $c$ closed to zero, confirming a thermodynamic vanishing. Crucially, this hexagon GMN extinction at criticality serves as a nonlocal order parameter for ferromagnetic ordering.

The entanglement architecture of non-Landau criticality is further exemplified by the Néel-VBS transition in the $H_{JQ_3}$ model with a pinning field. 
As shown in Fig.~\ref{fig:jq3GMN} (a), typical unicursal subregions show complete GMN extinction near criticality, except for the four-party GMN in the loopy plaquette. Importantly, this remaining GMN shows a local minimum at the QCP, which directly opposes the typical Landau-type peak in Fig.~\ref{fig:O2O3_GMN} (a,c). For non-unicursal geometries (Fig.~\ref{fig:jq3GMN} (b)), we find a finite three-party GMN in cross-shaped regions at QCP. This entanglement is fragile; reducing it to $T$-shaped subregions through single-site tracing destroys GMN. Interestingly, the five-party cross-shaped GMN disappears exactly at criticality, making it a Néel order parameter. 
In both the XY$^*$ and Néel-VBS transitions, we find that three-spin GMN remains undetected, and no GMN peak is observed at the QCP for any studied subregion. Instead, for certain subregions, GMN either vanishes or reaches a minimum at criticality. This contrasts fundamentally with the entanglement enhancement seen in Landau criticality. 
The suppression of GMN in small subregions implies a redistribution of critical entanglement to large subregions (beyond 6 spins) at non-Landau QCPs, where the entanglement architecture becomes more collective.

\begin{figure*}[htp!]
\includegraphics[width=\textwidth]{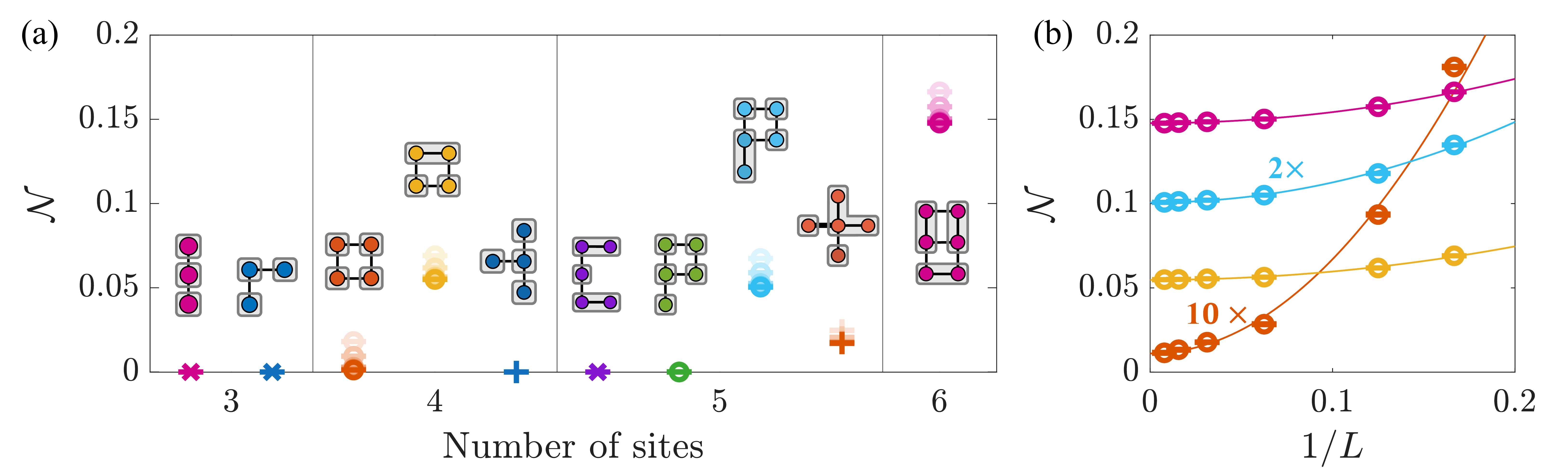}
\caption{\textbf{GMN for RVB wavefunction.} (a) The GMN for RVB wavefunction in different subregions. The color intensity of the data points ranges from light to dark, representing different system sizes: $L = 6,8,16,32,64,128$. Asterisk, circle, and plus-sign markers represent loopy, non-loopy, and non-unicursal subregions. 
(b) Finite-size extrapolation using a power law fitting, i.e, $aL^{-2}+c$ for loopy subregions with the same colors used in panel (a). Some GMN values are scaled up by a constant for clarity.}
\label{fig:RVB_GMN}
\end{figure*}

\newsect{GMN for RVB state}
Given the clear difference in entanglement architecture between Landau and non-Landau critical points, we extend our analysis to the paradigmatic spin-1/2 resonating valence bond (RVB) state on the square lattice~\cite{Liang_Doucot_Anderson_RVB_1988, Albuquerque_Alet_2010, Tang_Sandvik_Henley_2011, Zhang_Beach_2013, Torlai_Melko_2024}. This critical state exhibits power-law correlations governed by an emergent U(1) gauge theory~\cite{Wen1996UnderdopedCUprates,Patrick2006Doping}. The RVB wavefunction is defined as
$\ket{\psi_{\textrm{RVB}}} = \sum_{\mathcal{C}} \prod_{(i,j)\in \mathcal{C}} \ket{b_{ij}}$,
where the summation runs over all nearest-neighbor dimer coverings $\mathcal{C}$ of the lattice in the $(0,0)$ topological sector, with $\ket{b_{ij}} = \frac{1}{\sqrt{2}} \left( \ket{\uparrow_i\downarrow_j} - \ket{\downarrow_i\uparrow_j} \right)$ denoting the singlet bond between sites $i$ and $j$.

Fig.~\ref{fig:RVB_GMN} presents GMN results for RVB wavefunctions, computed using our generalized entanglement microscopy algorithm (see SM~\cite{suppl} for a detailed description), which are converged with system size to reflect thermodynamic behavior.
Notably, RVB entanglement reflects non-Landau criticality: three-spin GMN is absent, unicursal subregions show no non-loopy GMN, which agrees with previous studies with $L=6$ exact wavefunction~\cite{lyu2025loop}. The minimum subregion that detects GMN is the four-site plaquette region, although the four-party GMN within it extrapolates to a tiny value in the thermodynamic limit. 
We also found finite but fragile GMN in non-unicursal cross-shaped areas, which disappear when reduced to $T$-shaped geometries by single-site tracing. This exact correspondence with Néel-VBS criticality confirms that the observed entanglement architecture is a universal feature of deconfined gauge theories.

Lastly, we consider the minimal multipartite entangled
subregion (MMES), which denotes the smallest subregion that hosts GME. As summarized in Table~\ref{table:tab1}, the form of the MMES distinguishes between different types of QCPs and critical states studied in this work. Specifically, the MMES in conventional Landau-type QCPs is the smallest three-spin L-shaped region. In contrast, the MMES takes on extended, loopy regions at non-Landau transition points or in critical states with emergent gauge field descriptions. At the XY$^*$ QCP, the GMN for a hexagon region extrapolates to zero in the thermodynamic limit, suggesting that the actual MMES must be larger than a single hexagon. This implies that GME is not stored in such local units but requires even more extended, collective structures, underscoring the highly non-local character of the beyond-landau quantum criticality.

\begin{table}[h]
\caption{{Minimal multipartite entangled subregion (MMES) at the various transitions.}}
\begin{tabular}{c|c|c|c|c}
\hline
\hline
     & O(2)/O(3)              & XY$^*$                 & N\'eel-VBS               & RVB state             \\ \hline
MMES & \multicolumn{1}{c|}{\lshape} & \multicolumn{1}{c|}{larger than  \hexagonwithdots} & \multicolumn{1}{c|}{\plaquette} & \multicolumn{1}{c}{\plaquette} \\ \hline \hline
\end{tabular}
\label{table:tab1}
\end{table}

\newsect{Beyond multiparty negativity}{
Since a vanishing GMN does not preclude genuine multipartite entanglement, we certify the biseparable (BSEP) nature of states in the various models via the adaptive polytope method~\cite{Ties2024}. 
In the BFG model (Fig.~\ref{fig:BFG_GMN_all_size}), the bowtie subregion transitions from being BSEP in the QSL phase ($J_\pm/J_z\leq0.06$) to a GME state near the QCP where GMN first appears. The triangle subregion, however, remains separable (SEP) across the critical point ($J_\pm/J_z \leq 0.07$) and becomes only BSEP in the FM phase. For the hexagon subregion, although a weak GMN is still present at the QCP for $L=8$, we can show that by adding a small white noise of $p=0.06$, the state becomes BSEP.
Notably, these sharp transitions in entanglement architecture occur precisely around the QCP, underscoring the sensitivity of multipartite entanglement as a probe for phase transitions.
For the $JQ_3$ model (Fig.~\ref{fig:jq3GMN}), the four-spin linear subregion is BSEP precisely when the GMN vanishes ($q=0.66$), while the 5-spin cross subregion, despite its GMN vanishing near the transition, retains a small residual GME that only fully disappears at a larger $q \approx 0.9$, where it finally becomes BSEP.
Finally, in the RVB state (Fig.~\ref{fig:RVB_GMN}), all subregions with zero GMN are certified as BSEP, though none are fully SEP due to finite bipartite entanglement between nearest neighbors. 
Overall, the faithful correspondence between vanishing GMN and biseparability confirms GMN as a robust and practical witness for GME across these diverse systems.
}

\newsect{Conclusion and outlook}
We have established the spatial architecture of GME as a key feature that sets apart Landau from non-Landau quantum criticality. Through entanglement microscopy of the O(2), O(3), XY$^*$ and Néel-VBS transitions, as well as the RVB states, we discovered a universal distinction: Landau criticality \textit{magnifies} GME within minimal subregions, while 
non-Landau criticality \textit{redistributes} entanglement into larger, mainly loopy structures. This important distinction is based on the absence of three-spin GMN, the lack of non-loopy GMN in unicursal subregions, and the suppression of GMN values in small subregions near non-Landau QCPs. 
{
Our work establishes the power of entanglement microscopy to directly probe the fundamental reorganizations of quantum entanglement at a microscopic level, revealing emerging gauge structures and fractionalized excitations. In contrast to conventional measures based on entanglement entropy~\cite{Swingle2012Structure,Isakov2011Topological,Zhao2022Measuring}, which offer indirect inference, our technique achieves direct tomography of multipartite entanglement architecture, mapping its geometrical distribution across strongly correlated systems.
}

Looking ahead, promising directions may include: (i) Using GMN spatial patterns as fingerprints to identify non-Landau quantum critical points and precisely locate phase boundaries; (ii) Developing general principles to systematically understand and identify order-parameter-like entanglement features, such as the vanishing three-party hexagon GMN at XY$^*$ QCP and five-party cross-shaped GMN at N\'eel-VBS QCP, for various symmetry-breaking patterns and lattice geometries; (iii) Validating these signatures in quantum simulators~\cite{Bluvstein2022quantumprocessor,Bourennane2004Experimental,Zhang2023Scalable}. This framework opens a new approach where the multipartite entanglement structure serves as both a diagnostic tool and a basic description of quantum critical phenomena.

\newsect{Acknowledgment}
MHS, TTW and ZYM acknowledge the support from the Research Grants Council (RGC) of
Hong Kong (Project Nos. 17309822, C7037-22GF, 17302223, 17301924), the ANR/RGC Joint Research Scheme sponsored by RGC of Hong Kong and French National Research Agency (Project No. A\_HKU703/22). We thank HPC2021 system under the Information Technology Services at the University of Hong
Kong, as well as the Beijing Paratera Tech Corp., Ltd~\cite{paratera} for providing HPC resources that have contributed to the research results
reported within this paper. W.W.-K.\/ and L.L.\/ are supported by a grant from the Fondation Courtois, a Chair of the Institut Courtois, a Discovery Grant from NSERC, and a Canada Research Chair.

\bibliographystyle{longapsrev4-2}
\bibliography{bibtex}

\clearpage
\twocolumngrid

\begin{center}
	\textbf{\large Supplemental Material for \\"Entanglement architecture of beyond-Landau quantum criticality"}
\end{center}
\setcounter{equation}{0}
\setcounter{figure}{0}
\setcounter{table}{0}
\setcounter{page}{1}
\setcounter{section}{0}

\makeatletter
\renewcommand{\theequation}{S\arabic{equation}}
\renewcommand{\thefigure}{S\arabic{figure}}
\setcounter{secnumdepth}{3}

\section{RDM sampling for the RVB state} \label{Appendix:RVB_RDM}
The updating scheme is based on the combined spin-bond basis updated from Ref.~\cite{Sandvik_VB_PQMC_2010}. With spin configuration explicitly defined throughout the process, the simulation can be easily adapted to compute the reduced density matrix for any subregion, albeit the number of matrix elements scales exponentially with the number of sites in that region.

The RVB wave function
$$
\ket{\psi_\textrm{RVB}}=\sum_{C}\prod_{(i,j)\in C}\ket{b_{i,j}}
$$
with $\ket{b_{i,j}}=\frac{\sqrt{2}}{2}\left (\ket{\uparrow_i\downarrow_j}-\ket{\downarrow_i\uparrow_j}\right )$. The first summation is taken over all possible dimer configurations $C$ containing nearest-neighbor pairs $\left (i,j\right )$ covering every site on the square lattice exactly once. On a bipartite lattice like the square lattice, one can assume, without loss of generality, that all sites $i$ ($j$) are on sublattice $A$ ($B$).

We first rewrite the wave function to the spin basis. For a given dimer configuration C, one can expand the sum on each singlet state and obtain
$$\ket{\psi_C}=\prod_{(i,j)\in C}\ket{b_{i,j}}\propto\sum_{\{\sigma^z\}\cap C}(-1)^{N_{A,\downarrow}(\{\sigma^z\})}\ket{\{\sigma^z\}}.$$
That is, we go back to using the spin basis ($\sigma^z=\uparrow,\downarrow$) on each site, and only sum over spin configuration that are compatible with the dimer configuration C, i.e., the spin orientation on sites $\left (i,j\right )$ are always opposite to each other. ${N_{A,\downarrow}(\{\sigma^z\})}$ counts the number of spins down on sublattice A, and the phase $(-1)^{N_{A,\downarrow}(\{\sigma^z\})}$ originates from the minus sign in the singlet state.

On this basis, the reduced density matrix for this state on a subregion $X$ is simply
\begin{align*} 
\rho_X&=\Tr_{\bar{X}} \left (\ket{\psi_\textrm{RVB}}\bra{\psi_\textrm{RVB}} \right ) \\ 
&\propto \sum_{C_1,C_2}\sum_{\{\sigma^z\}_1\cap C_1}\sum_{\{\sigma^z\}_2\cap C_2} \\
&(-1)^{N_{A,\downarrow}(\{\sigma^z\}_1)+N_{A,\downarrow}(\{\sigma^z\}_2)} \Tr_{\bar{X}} \left (\ket{\{\sigma^z\}_1}\bra{\{\sigma^z\}_2}\right ),
\end{align*}
where we need two replicas with spin orientations $\{\sigma^z\}_1$ and $\{\sigma^z\}_2$, and they must be compatible with their corresponding dimer configurations $C_1$ and $C_2$, respectively. The partial trace on the complement region $\bar{X}$ enforces that the spin configurations inside $\bar{X}$ are the same on the two replicas.

\begin{figure*}[htp!]
\includegraphics[width=2\columnwidth]{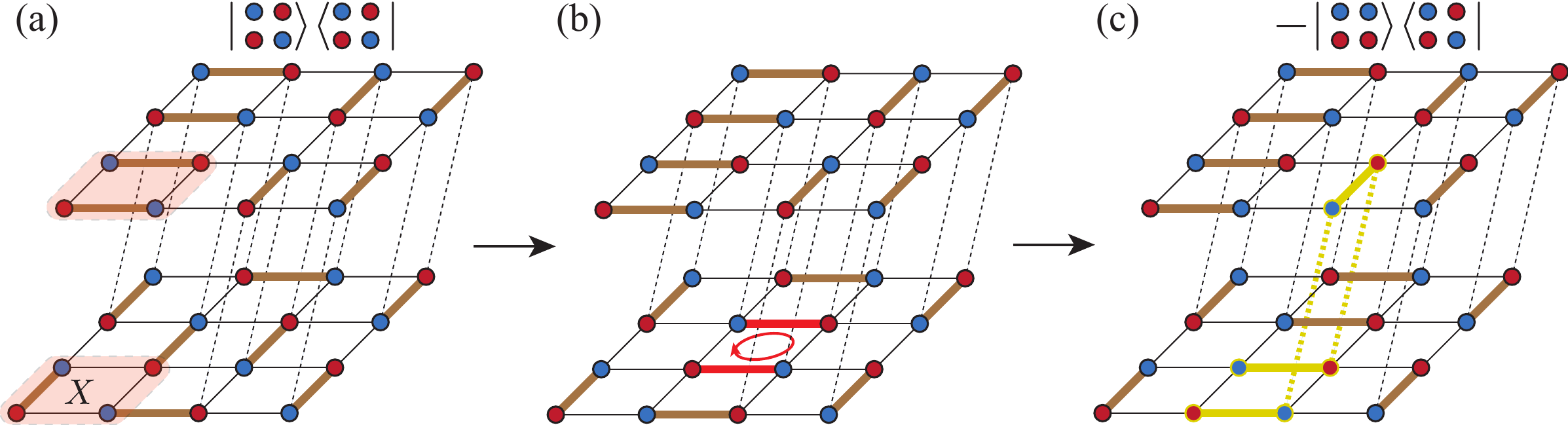}
\caption{\textbf{Schematic of sampling the reduced density matrix for RVB state.} In all panels, the spins in region $\bar{X}$ are the same for the two replicas (connected by dotted lines), and the thick brown bonds represent the singlets. Circles in red and blue are the spin up and down on every site. (a) The initial state of the configurations, with the corresponding RDM element on top. (b) The plaquette in the middle is chosen to update, and the red bonds are rotated from being vertical to horizontal. (c) A cluster of spins (solid and dotted path in yellow) is chosen to be flipped. The RDM element sampled in this updated configuration is on top, and the minus sign comes from the one flipped spin from sublattice $A$ in region $X$.}
\label{fig:RVB_RDM}
\end{figure*}

Fig.~\ref{fig:RVB_RDM} (a) shows an example of a $4\times 4$ lattice. The thick brown bonds represent the dimers, and the red and blue circles are spin up and down on each site. The four sites at the lower left corner (red shaded square) are chosen to be the subregion $X$, and all the other sites in $\bar{X}$ from the two replicas are connected by dotted lines because of the partial tracing condition.

The updating scheme consists of two steps: bond update and spin update. For the bond update, a plaquette is picked randomly. Then it is flipped if there are parallel dimers around this plaquette and the spin configuration is compatible with the new dimer configuration. This process is repeated multiple times for both replicas separately. As shown in Fig.~\ref{fig:RVB_RDM} (b), the plaquette in the center is chosen and flipped from being vertical to horizontal (red bonds in Fig.~\ref{fig:RVB_RDM} (b)) since it fulfills both conditions.

For the spin updates, clusters are grown around a randomly chosen site, and all spins on the clusters are flipped. A cluster consists of sites that are linked together either due to the partial tracing condition (dotted line in Fig.~\ref{fig:RVB_RDM}) or by dimers (thicker bonds in Fig.~\ref{fig:RVB_RDM}). As shown in Fig.~\ref{fig:RVB_RDM} (c), the yellow bonds and dotted lines constitute a cluster, and all the spins on it are flipped.

For the case of taking a full trace, this reverts to the original sampling with a combined basis, where there is only one spin configuration, and the overlaying of two bond configurations always forms closed loops on the lattice.

\section{Genuine Multipartite Negativity as a Semidefinite Program}
The Genuine Multipartite Negativity (GMN) defined in Eq.(\ref{eq:CRE_GMN}) can be computed via the following semidefinite program (SDP)\cite{Guhne2011, Hofmann_2014}:
\begin{equation*}
\begin{aligned}
\mathcal{N}_{\text{GM}}(\rho) &= -\min \operatorname{tr}(\rho W) \\
\text{subject to } &W = P_m + Q_m^{T_m}, \\
& 0 \le P_m, \quad 0 \le Q_m \le I \quad  \\
& \text{for all bipartitions } m \mid \bar{m},
\end{aligned}
\end{equation*}
where W is an entanglement witness which is fully decomposable along any bipartition, and $T_m$ denotes the partial transpose over one side of the bipartition. The corresponding dual problem is~\cite{Hofmann_2014} 
\begin{align*}
 \min p_A &\operatorname{tr}\left(\rho_A^{-}\right)+p_B \operatorname{tr}\left(\rho_B^{-}\right)+p_C \operatorname{tr}\left(\rho_C^{-}\right) \\
\text {such that } \rho&=p_A \rho_A+p_B \rho_B+p_C \rho_C 
\\
\quad \rho_m^{T_m}&=\rho_m^{+}-\rho_m^{-} \\ \text {for all } m &\in\{A, B, C\} \text { with } \rho^{ \pm}_m,\rho_m \geq 0,
\end{align*}
This suggests an operational definition: GMN detects multipartite entangled states that are outside a mixture of states that is larger than the biseparable set. GMN detects states that lie outside the set of \textbf{PPT-mixtures}, which is the convex hull of states that are PPT with respect to some bipartition. In the tripartite case, this set takes the form:
\begin{equation*}
    \rho^{\mathrm{pmix}} = p_1 \rho_{A \mid BC}^{\mathrm{ppt}} + p_2 \rho_{B \mid AC}^{\mathrm{ppt}} + p_3 \rho_{C \mid AB}^{\mathrm{ppt}}.
\end{equation*}
All biseparable states are contained within this set; they are mixtures of states that are separable across some bipartition:
\begin{equation*}
    \rho^{\mathrm{bs}} = p_1 \rho_{A \mid BC}^{\mathrm{sep}} + p_2 \rho_{B \mid AC}^{\mathrm{sep}} + p_3 \rho_{C \mid AB}^{\mathrm{sep}}.
\end{equation*}
By construction, \(\mathcal{N}_{\text{GM}}(\rho) \ge 0\), and the convexity of the formulation ensures global optimality of the solution.
If a state lies outside the PPT mixture set, the SDP yields a finite value of \(\mathcal{N}_{\text{GM}}(\rho)\), certifying genuine multipartite entanglement. If the value is zero, the result is inconclusive. 
\section{Extra data for BFG model}
\subsection{Temperature convergence}

\begin{figure}[htp!]
\includegraphics[width=\columnwidth]{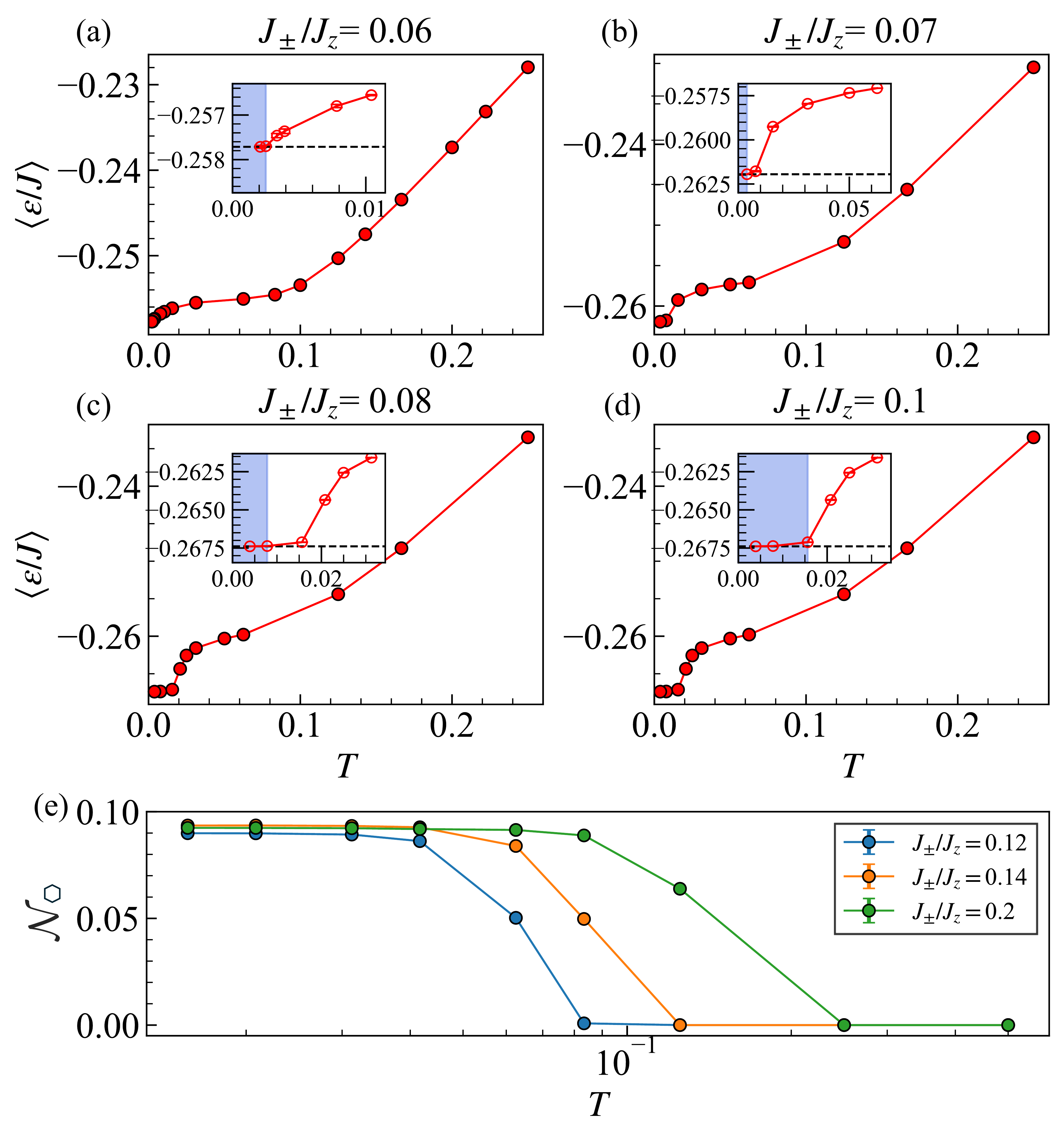}
\caption{\textbf{Temperature convergence of BFG model at $L=6$.} Panels (a)-(d) show the energy per site ${\varepsilon}=E/N$ against temperature at various $J_{\pm}/J_z$ values. The shaded areas in the insets present the temperature region where the ground state property is well preserved. Panel (e) shows the evolution of three-party GMN in a hexagon region against temperature at three $J_{\pm}/J_z$ values inside the ferromagnetic phase.}
\label{fig:T_convergence}
\end{figure}

Fig.~\ref{fig:T_convergence}(a)-(d) shows how the energy per site converges with temperature at typical coupling strengths exemplified at $L=6$. Inside the spin liquid phase (panel (a)), two energy plateaus appear. The higher-temperature plateau corresponds to the spinon gap energy scale ($\Delta_s$). The lower-temperature plateau corresponds to the smaller vison gap ($\Delta_v$). Therefore, probing the true ground state requires very low temperatures, within the shaded area shown in the insets. Near the quantum critical point (QCP) and inside the ferromagnetic (FM) phase, these plateaus become less distinct (though finite-size effects may cause remnants) and the convergence temperature rises. We compute reduced density matrices only within the blue shaded temperature region to avoid thermal effects. Panel (e) tracks the three-party genuine multipartite negativity (GMN) in a hexagon region, $\mathcal{N}_{\hexagon}$, with temperature in the FM phase. $\mathcal{N}_{\hexagon}$ converges at low temperatures but abruptly vanishes above a threshold temperature, i.e., \textit{sudden-death} temperature, remaining zero at higher temperatures.
\subsection{Exact diagonalization at $L=3$}

\begin{figure}[htp!]
\includegraphics[width=0.8\columnwidth]{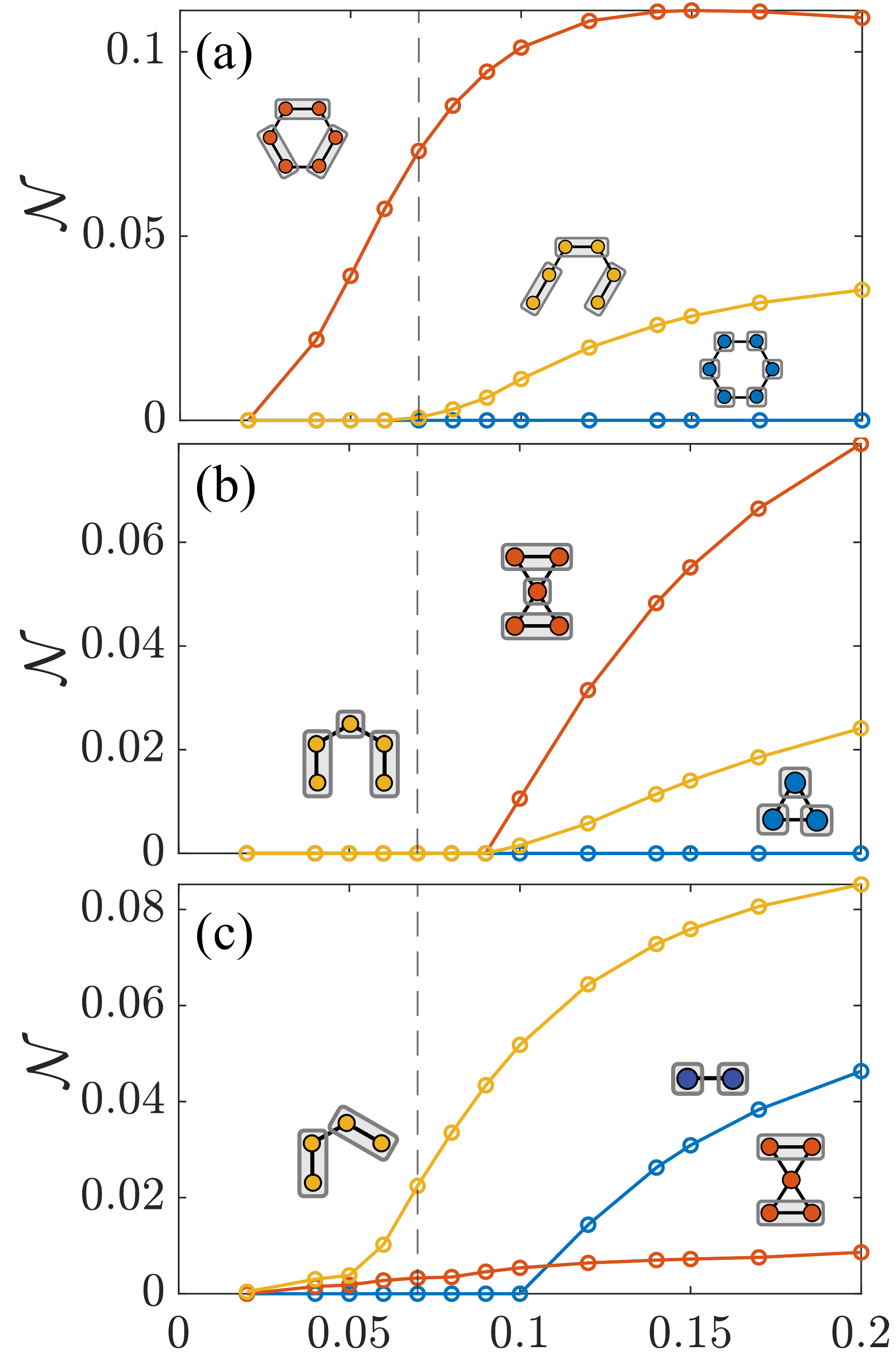}
\caption{\textbf{GMN of BFG model from exact diagonalization.} Panels (a)-(c) shows the GMN results of BFG model at $L=3$ obtained by exact diagonalization with various subregions and partitions. The dashed line represents the location of QCP at $J_{\pm}/J_z\approx 0.07$.}
\label{fig:ED_BFG}
\end{figure}

We compute the GMN via exact diagonalization (ED) for the $L=3$ BFG model (27 sites); results for representative subregions and partitions are shown in Fig.~\ref{fig:ED_BFG}. These ED results agree qualitatively with the QMC data presented in the main text, which were obtained for larger systems. The strong finite-size effects—particularly near the QCP and within the $\mathbb{Z}_2$ QSL phase—are observed, which results in a vanishing GMN for all subregions studied at larger system sizes.

\section{Effect of pinning field on $JQ_3$ model}

\begin{figure}[htp!]
\includegraphics[width=0.8\columnwidth]{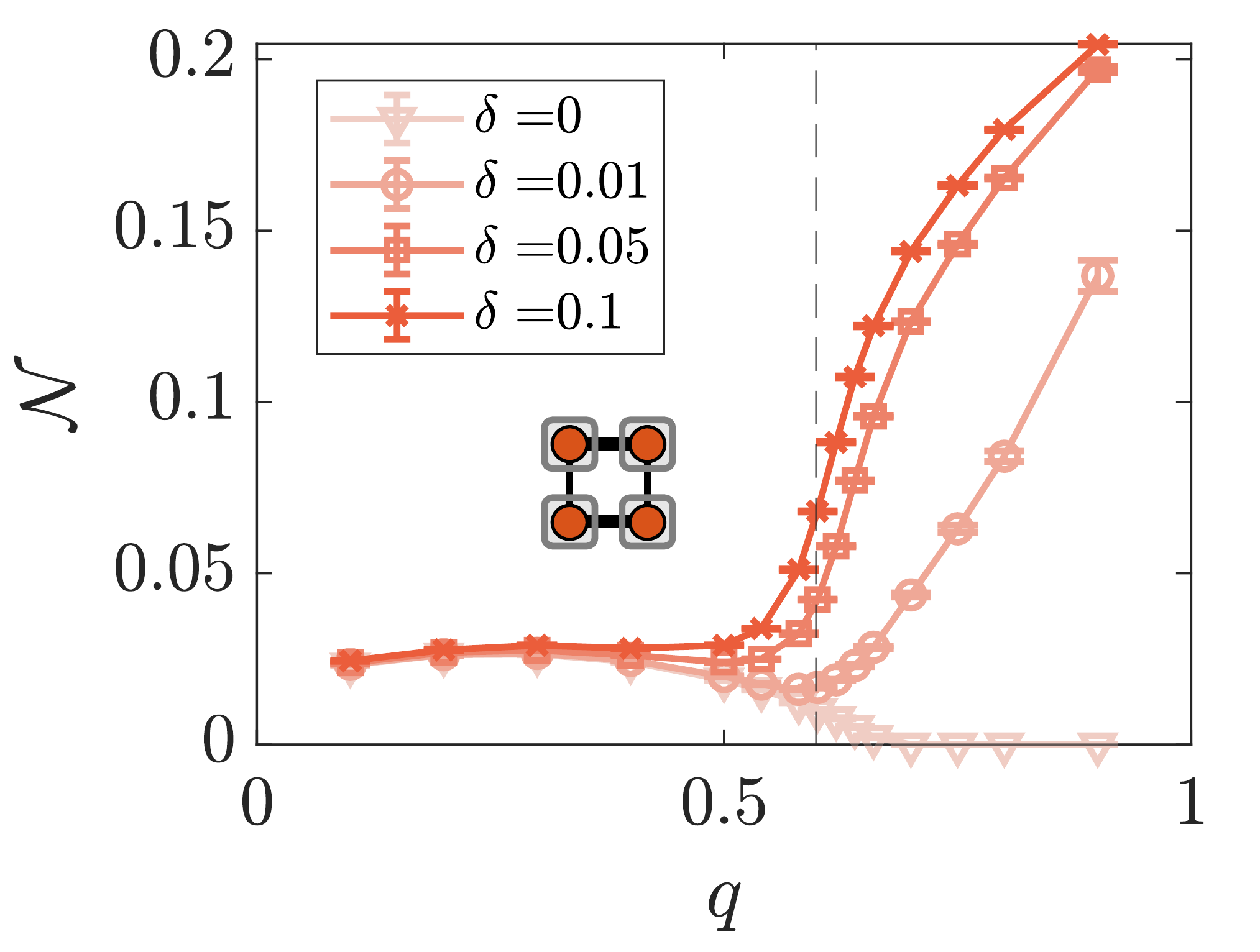}
\caption{\textbf{Effect of pinning field on GMN in $JQ_3$ model.} The four-party GMN in the plaquette region is chosen to demonstrate the effect of pinning field with strength $\delta/L$ at $L=24$ and inverse temperature $\beta=96$.}
\label{fig:pinning_field_JQ}
\end{figure}

As mentioned in the main text, we add a tiny pinning field of strength $\delta/L$ to $H_{JQ_3}$ at the $J_2$ bonds defined in Fig.~\ref{fig:fig1} (a). This pinning field is expected to lift the fourfold degeneracy of the valence bond solid (VBS) phase while keeping the critical behavior intact. At infinite temperature, an infinitesimal pinning field, in principle, can pick out one VBS configuration out of the four degenerate ones. While given a fixed system size and temperature, the pinning field needs to exceed a threshold such that $\rho=e^{-\beta H}$ gives a pure state density matrix. Fig.~\ref{fig:pinning_field_JQ} demonstrates the effect of the pinning field on the four-party GMN in the plaquette region at a given system size $L=24$ and temperature $\beta=1/T=96$. The GMN varies smoothly with coupling $q$ and vanishes inside the VBS phase, showing no peak or singularity near the quantum critical point (QCP). Adding a pinning field of $\delta=0.01$ increases the GMN within the VBS phases but leaves the GMN at the QCP and inside the Néel phase unchanged. Larger pinning fields do affect the GMN at the QCP and in the Néel phase, indicating the field is too strong for this $L$ and $\beta$. Therefore, our simulations use $\delta=0.01$ and $\beta=96$ at $L=24$ to lift the VBS degeneracy while protecting criticality.

\end{document}